\documentstyle{article}
\begin{document}
\begin{center} Spec. in Sci. and Tech. {\bf 17}, 232-36 (1994) r
\end{center}
\begin{center}{\Large{\bf Life and Water}\\

H.C. Rosu
\footnote{e-mail: rosu@ifug.ugto.mx

Additional address: IGSS, Magurele-Bucharest, Romania}\\[2mm]
{\scriptsize{Instituto de F\'{\i}sica, Universidad de Guanajuato,\\
Apdo Postal E-143, 37150 Le\'on, Gto, M\'exico}\\  } }
\end{center}


\begin{center}
{\bf Abstract}

Many living organisms on Earth are strongly dependent on water,
the natural liquid of the planet. A possible reason for that could be
the conjecture of Ryoji Takahashi [Phys. Lett. {\bf A 141}, 15 (1989)]
that water microdrops release
negentropy through a phase transition to a phase with zero surface
tension. Biological cells could make use of such a phase transition in
their duty cycle.
We comment on the relative merit of this conjecture, and present
it in wider theoretical context.
\end{center}


\vskip 0.3cm

IC-246/1992 r [$\cal H \cal C \cal R$] \hfill LAA physics/9610013

\vskip 0.5cm




\section{Introduction}

From time to time physicists are venturing in realms not very familiar
to them, like biology (see two famous examples in the following).
 According to Dyson, the {\it noblesse oblige} principle
of not writing on topics in which the author is not a master prevented
many from such activities in the past. But topics of extremely broad
character, like the origins of life, are literally open to everybody as
soon as one knows how to ask the right questions.

In February 1943, Erwin Schr\"{o}dinger gave a famous course
 of lectures to an
audience of about four hundred at Trinity College, Dublin. One year
later this course was published as the little book {\it What is Life ?}.
 It is an influential book to this day and a masterpiece, read
by all the creators of molecular biology in the fifties as well as by
many more unknown people. Schr\"{o}dinger's
book is essentially describing the phenomenon of biological replication,
and only in one chapter the other important life phenomenon of
metabolism. Also the concept of negentropy is introduced.

In February 1985, Freeman Dyson delivered the Turner Lectures  at
Trinity College, Cambridge, published as a little book with the title
{\it Origins of Life}. One will find in this book a discussion of the
so-called double-origin hypothesis of life. The first beginning of life
is with replication (protein) creatures and the second beginning is with
creatures capable of metabolism.

We recall here the astroplankton hypothesis of Haldane$^{\cite{0}}$,
i.e., vital units wandering through the universe, ready to start
the cycle of evolution whenever conditions are favorable. The same
author$^{\cite{01}}$ gave an estimate for the spontaneous occurence of
a bacterial virus in the primitive ocean. Such a creature has,
according to Haldane, about 100 bits of negative entropy of
information that could arise spontaneously in the primitive water
in $10^{9}$ years.

The most remarkable chemical characteristic of living matter is its
optical activity. For recent progress the reader is referred to the
Ciba symposium$^{\cite{02}}$.
 What we call living organisms synthesize themselves
from molecular subunits of one enantiomorph alone. This surprising fact
was already known to Pasteur. In his lectures, as early as 1860, he
popularized the idea that biosyntheses involve a chiral force. After
the discovery of parity violation in weak interactions (1956),
and especially
after people became aware of the electroweak unification scheme (1970),
there were speculations tracing the origin of biochiral forces to the
electroweak parity violation$^{ \cite{03}}$. However, it has
 to be clearly stated
that electroweak energy difference between the enantiomers of an
amino acid corresponds to an excess of $10^{6}$ molecules of the stable
form in a total of $10^{23}$, i.e., one part in $10^{17}$. This is only
an extremely weak ``signal" onto the background of many random
 fluctuations in more important factors, like circularly polarized
 light, that could determine which enantiomer is formed. However,
 some time ago, Salam$^{\cite{sa}}$ proposed an enhancement
 mechanism of the
 chiral electroweak signal by means of a phase transition into a
 condensed Bose-like mode. See also the comments of
 Chela-Flores$^{ \cite{cf}}$, and the recent work of
  Salam$^{ \cite{s}}$.

In the widely interdisciplinary problem of the origins of life it is
very difficult to step out from the domain of speculations into the
domain of firmly established experiments.

In the following we shall discuss a new point of view on the
strong correlations among the life forms on Earth and the water liquid.
It is based on the works of Ryoji Takahashi$^{ \cite{1}}$ on the
negentropy of micron size drops of water. This negentropy is related
to the surface tension coefficient of water drops. Laplace's equation
and Maxwell's third thermodynamic relation are the key features in
stating the negentropy property of water microdrops when their
volume is increased. To explain thermodynamically
the liquid phenomena in the micron region, which have been observed by
reflected scanning electron microscopy (SEM), Takahashi introduced
the hypothesis of a new phase of water, the super water. The scope of
this short paper is to comment on the results of Takahashi.

\section{Normal water}

At the molecular level, water is an extremely unusual substance. It
has two types of intermolecular forces - the hydrogen bond and the
hydrophobic effect. Although of low molecular weight, water has
 unexpectedly high melting and boiling points and latent heat of
 vaporization. Besides, liquid water exhibits the famous density
 maximum at $4^{o}C$, and it is hevier than its solid phase (ice).
 Usually one assumes that the strong intermolecular bonds formed
 in ice persist into the liquid state and that they must be strongly
 orientation-dependent since water adopts a tetrahedral coordination;
 more exactly the average number of nearest neighbors is 4.5. For more
 details on the cluster structure of water the interested reader
 may consult the recent paper of Benson and Siebert$^{ \cite{bs}}$.

Other unusual properties are the very low compressibility and
curious solubility features, both as a solute and as a solvent.
Liquid water possesses high molecular dipole moment and high dielectric
constant, which, unlike any typical polar liquid, is actually
increasing when water freezes into ice, and is still increasing at
$-70^{o}C$.

\section{Super water}

The surface tension of microdrops of water was investigated by means of
reflected scanning electron microscopy (SEM) and
very interesting phenomena were reported ten years ago by
 Takahashi$^{\cite{t1}}$. He investigated the creation of microdrops
 by the dehydration of hydrated metaphosphoric acid used as
 specimen when the surface of the specimen underwent bombardment of
 the electron beam. The imaging current was as low as $10^{-10}A$
 with a beam diameter less than $100 \AA$
 and the voltage was lowered in steps down to 5 kV. The irradiation
 of the surface was done for 3 to 15 s. At high voltages the
 surface developes craters, but at low voltages the change of the
 surface is a local rise in the form of a microdrop without any crater.
 This means that the beam ionizes locally the included water and
the pressure is increased in a narrow region under the surface,
giving birth to a microdrop according to Laplace's formula. In other
words, the SEM experimental conditions are just at the threshold of
molecular surface tension phenomena.
 Since the surface tension is due to attractive forces
between molecules acting over tens of $\AA$, the surface tension
concept is still in its range of validity for drops of micron sizes.
At low values of the voltage the ionization is diminished and
consequently the droplets should increase continuously in size.
Surprisingly, the experimental result is different. At low
accelerating voltages (from 5 kV to 8 kV) the size of the droplet
is almost independent of the voltage and of the size of the specimen.
The interpretation of Takahashi is to postulate
the existence of a minimum
drop in any liquid, a fact he claims to be universal. Moreover,
by employing Gibbs-Helmholtz equation he claims that in the one micron
region there exists a phase transition to a super- lubrifiant
phase of water for which the surface tension is naught.
The Gibbs-Helmholtz equation gives the surface energy of a plane liquid
 per unit area u as follows:
 $$\frac{u}{T} =\frac{\gamma}{T}-\frac{d\gamma}{dT}   \eqno(1)  $$
 or since the second term is just the surface entropy per unit area
 changed in sign:
 $$\frac{u}{T}=\frac{\gamma}{T} + s    \eqno(2)  $$
 A zero $\gamma$ implies both $u$ and $s$ zero. Such a limiting case
 describes a liquid which is able to freely deform itself without
 any change in volume, i.e., the best lubricant. Takahashi estimated the
 minimum diameter of water droplets at which the normal-super water
 transition
 takes place to be $1 \mu m$. As the short range attractive forces drop
 to naught in the super phase the binding is provided by the
  electromagnetic attractive forces, being less than that of normal
   water. In the surface of the super water the electromagnetic forces
   exerted on charged particles are balanced at each point in the
    tangent plane.

It is this normal-super water transformation that was thought
 responsible
by Takahashi of the negentropy cycle of living cells. These living
devices accumulate entropy by metabolic processes. The excess entropy
is however compensated by the negentropy of a pressure induced
 transformation to super water. A conclusion to be drawn from this
 explanation is
that the size of a one-cell organism should be larger than the
minimum diameter of normal-super water transition.

\section{Negentropy of water microdrops}

In 1989 Takahashi$^{ \cite{1}}$ calculated the Laplace pressure in water
 droplets of various sizes in the micron range in saturated moisture
based on the Landolt-B\"{o}rnstein data of $\gamma $ of 1956.
According to Takahashi,
the P-T curves represent the equilibrium between the normal and
the super phases. Within equilibrium conditions
Maxwell's third thermodynamic relation is fulfilled:
$$(\frac{\partial S}{\partial V})_{T}=(\frac{\partial P_{L}}{\partial T}
)_{V}     \eqno(3) $$
$P_{L}$ is the Laplace pressure. At small diameters and for room
 temperatures, $(\partial P_{L}/\partial T)_{V}$ is negative. Thus,
from Maxwell's relation, it follows that when normal water turns
into super water a negative entropy is produced in the transition.
This negative entropy of the normal-super water transition could be used
by biological cells in their attempt to avoid the second law of
thermodynamics. Another requirement is to create a negative
increment of pressure in the cell in order to drive the phase transition
to super water. The cell might do that by controlling electrically
the osmotic pressure. In the p-V plane this results in a pressure
induced transformation in which a work is given out.
 This clearly requires entropy. The argument of Takahashi is that the
biological cell produces, by metabolism, just the amount of entropy
required to compensate the negentropy of super water transformation.

\section{Conclusions}

 We presented some comments on the Takahashi
model of the so-called normal-super phase transition in water
 microdrops and
its application to the engine biological cell - microdrop. As soon as
one assumes the existence of a normal to super water phase
 transformations with a negative latent heat, one may consider the
 problem of the related negative entropy (negentropy). This is what
 Takahashi has done.
Even though the model looks very ambiguous, it is apparently supported
 by some
experimental evidence, and in spite of the fact that other explanations
 of the SEM data are possible,
it gives a rather challenging interpretation of the Schr\"{o}dinger
negentropy.

We would like to emphasize the fact that from the chemical
point of view cells are soup bubbles made of a great number of
substances. Water is the most abundant substance in biosystems,
at the level of 70\%, but actually one will find it almost entirely
in the form of intercellular water. A model for water in biosystems,
based on the assumption that water is adsorbed on macromolecules
has been presented by Cerofolini$^{ \cite{cer}}$.

Concerning Takahashi's hypothesis, we draw attention to the recent
work of Landsberg and Woodward$^{ \cite{lw}}$. When discussing the
incremental heat input into a thermodynamical system, they introduce
a ``latent heat'' $l_{X}$ corresponding to some function of pressure and
volume X(p,V) for which $l_{X}dX$ is not generally the work done on
the system. We are currently investigating the application of such an idea
to water microdrops in biological environments.

\section{Note added on August 5/1996}
I maintained a sporradic interest in Takahashi's experiment on
{\em minimum drop},
both on the negative and positive side. On the prevailing negative one,
I have to mention that Takahashi asseverated that
``measurement of the minimum drop with
any material other than metaphosphoric acid is nearly impossible"
and therefore the existence of a {\em universal minimum drop} in any liquid
is only a speculation.

About a year ago, I have devoted three days to this old file and I came up
with the idea that most probably Takahashi got into a miscibility
(and {\em meniscus} formation) problem, but I was not able to find
around any data on hydrated $HPO_3$ and
I let it pass away, since at that time I got involved in many other projects.
Even in this case the problem is really interesting and I offer
some references to those wishing to pursue such a study \cite{misc}.

\section*{Acknowledgements}

We thank Professor Abdus Salam, the International
Atomic Energy Agency and the United Nations Educational,
Scientific and Cultural Organizations for their hospitality
at the International Centre for Theoretical Physics in Trieste.



\end{document}